\documentclass[runningheads]{llncs}
\usepackage[T1]{fontenc}
\usepackage{url,hyperref}
\usepackage{amsmath,amssymb}
\usepackage[table]{xcolor}
\usepackage{multirow}
\usepackage{comment}
\usepackage{amscd,amsmath}

\usepackage{graphicx}
\usepackage{color}

\begin{document}
\title{A New Approach in Cryptanalysis Through Combinatorial Equivalence of Cryptosystems}
\titlerunning{Cryptanalysis Through Combinatorial Equivalence}
%
\author{Jaagup Sepp\inst{1} \and Eric Filiol\inst{2}}
	
\authorrunning{J. Sepp \& E. Filiol}
%
\institute{Hope4Sec, Crypto Lab, Tallinn, Estonia \\ \email{jaagup.sepp@hope4sec.eu} \\ \url{https://hope4sec.eu/}\and
	Independent Researcher - Retired Professor, Paris, France\\
	\email{efll@protonmail.com}\\ \url{https://ericfiliol.site}}

\maketitle              
\begin{abstract}
We propose a new approach in cryptanalysis based on an evolution of the concept of \textit{Combinatorial Equivalence}. The aim is to rewrite a cryptosystem under a combinatorially equivalent form in order to make appear new properties that are more strongly discriminating the secret key used during encryption. We successfully applied this approach to the most secure stream ciphers category nowadays. We first define a concept cipher called Cipherbent6 that capture most of the difficulty of stream cipher cryptanalysis. We significantly outperformed all known cryptanalysis. We applied this approach to the Achterbahn cipher and we obtained again far better cryptanalysis results. 

\end{abstract}

\section{Introduction}
\begin{figure}
	\centering
	\includegraphics[width=0.90\linewidth]{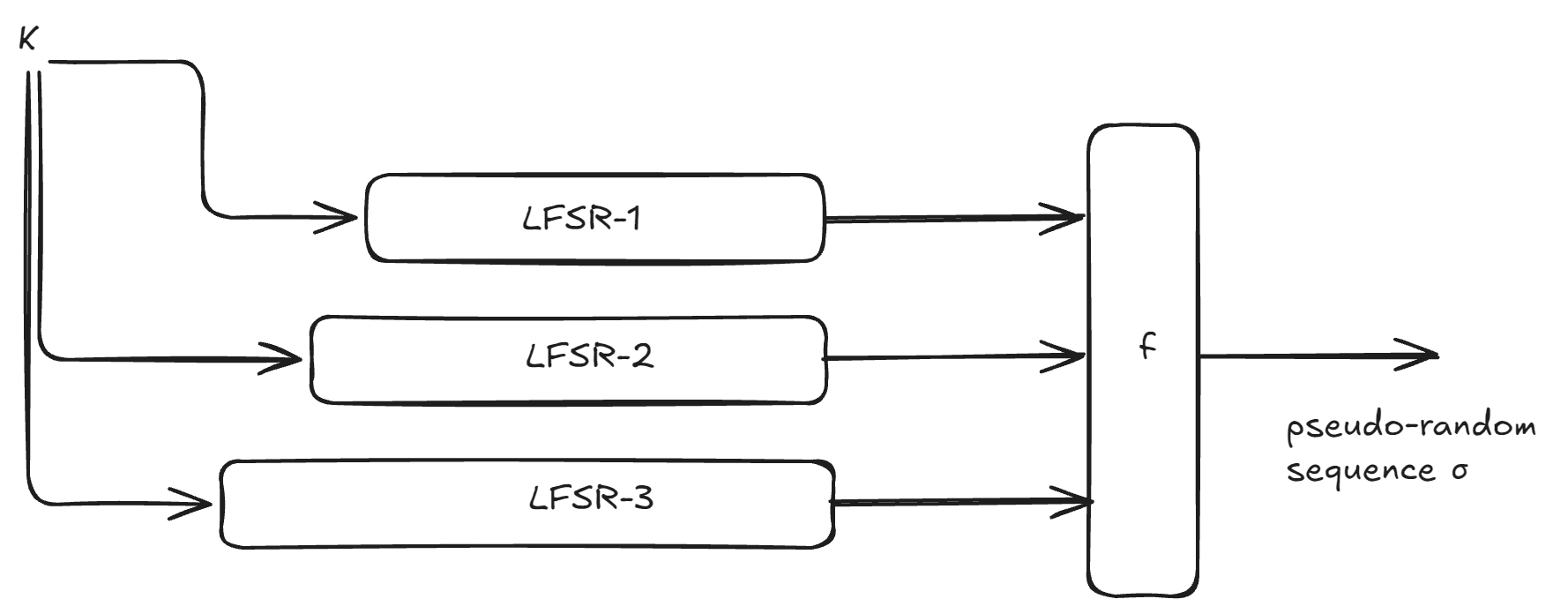}
	\caption{General Structure of LFSR-based Combiner} \label{fig1}
\end{figure}

Stream ciphers~\cite{Rueppel1986} are among the main cryptographic primitives used in symmetric cryptography. At the origin, the first stream ciphers were built with Linear Feedback Shift Registers (LFSRs), where linearity is meant in the register update function while the combining function is meant to be non-linear to break the intrinsic linear properties of the sequences produced by the register. The most common design is the combiner generator, depicted in Figure~\ref{fig1} to which most of other the stream cipher designs can be reduced.
Most stream ciphers use combined or non-linearly filtered LFSRs~\cite{Rueppel1986}. However, their security has been questioned over the years.
The natural evolution of these systems is towards the use of Non-Linear Feedback Shift Registers (NLFSRs)~\cite{Golomb1981} and Boolean functions with good cryptographic properties according to the general structure in Figure~\ref{fig1}. If for the latter a substantial body of knowledge exists~\cite{Carlet2021} -- with however a large number of problems still open as soon as the number of variables exceeds ten -- the study of NLFSRs is still in its infancy. To date, for example, no NLFSRs are known in maximum period for lengths greater than 31~\cite{nlfsr2014} when looking for rather simple, sparse feedback polynomials (a desirable property when dealing with implementation aspects). But the clever combination of these registers can lead to systems with effective security. The two best examples are Achterbahn~\cite{Acht2006} and Trivium~\cite{trivium}.

Attacking this new class of systems remains an open problem~\cite{Yao2021}. For real systems, no effective cryptanalysis is known that could seriously question their security. Algebraic or statistical attacks by correlation are no longer applicable. It is therefore necessary to consider a radically different approach that is neither algebraic nor statistical. 

We are working on such an approach applied to symmetric encryption (stream cipher and block ciphers). It is combinatorial in nature. The principle is to rewrite the system to produce a combinatorially equivalent system and to translate the key search on a ``more tractable'' instance, with a complexity that is reachable in practice (data and time) and lower than the complexity of known attacks in the algebraic or statistical domains. We have named this approach CE (\textit{Combinatorial Equivalence}). Note that the principle of rewriting in the algebraic domain was initiated in~\cite{Yao2021} for some specific ciphers but does not lead to significant gains in cryptanalysis complexity. The CE cryptanalysis applies both to stream and block ciphers. In this paper we focus on stream ciphers. The rewriting techniques just do not consider the same underlying combinatorial objects. It is also worth mentioning that the CE technique is fully transposable to LFSR-based stream ciphers linear stream ciphers and to ciphertext-only attacks.

We consider real-life cryptanalysis only. We denote it\textit{Effective Cryptanalysis}.
\begin{definition}[Effective Cryptanalysis]
	A cryptanalysis is called \textit{effective} whenever it can be performed: 
	\begin{itemize}
		\item In a limited time that allows the cryptanalysis to be played a finite number of times.
		\item On realistic input data sizes that are compatible with the operational reality of the use cases.
	\end{itemize}
\end{definition}
For instance, any attack requiring data without changing keys far beyond the crypto-periods in use in the industrial or governmental 
world is not effective. Moreover known plaintext attack are realistic as long as the amount of data does not exceed a few kilobits. Any attack whose time complexity is greater than $2^{70}$ is not effective nowadays. However, to evaluate the effective security of a system, the size of the required input data $N$  takes precedence over the time complexity $T$: a system which can be broken with a relatively small value of $N$ despite a high complexity $T$ is in practice considered less secure than a system requiring a much higher value of $N$ even for a significantly lower complexity. It is not so much impossible to increase computing power as it is to have enough input data (at least conceptually) at least if standard cryptographic policies are applied (in particular concerning crypto-periods).  

The paper is organized as follows. Section~\ref{comb-eq} formalizes and present the concept of Combinatorial Equivalence. Then in Section~\ref{sec:sota} the state-of-the-art with respect to stream ciphers cryptanalysis is summarized. Section\ref{sec:algospecs} presents the specification of the concept cipher \textit{Cipherbent6}. The intent is to assess the security of Cipherbent6 with respect to these known techniques. Subsequently, Section~\ref{sec:formal} summarizes the initial cryptanalysis results and performances we have obtained on Cipherbent6 with respect to the CE cryptanalysis. Finally, Section~\ref{sec:conc} presents the future work to develop the CE cryptanalysis.

\textbf{Disclaimer}: \textit{this paper is not basically a research article in the usual sense. The CE technique is not public and is reserved for the industrial world. This paper is intended to present our results. Their validity can be assessed on request by a challenge approach (sending an output sequence and returning the key).\\
Any views, opinions and materials presented in this paper are personal and are the results of the authors' own research work. They belong solely
to the authors and, in any case, they do not represent those of people, institutions, companies or organizations that the authors may or may not be associated with in professional or personal capacity (including past, present and future employers).
} 
\section{Combinatorial Equivalence of Cryptosystems} \label{comb-eq}
\subsection{Formal Definition of Cryptosystems}
We start from the initial definition of encryption schemes (or cryptosystems) given in \cite[Section 3.1.1]{Buchmann2004} that we have generalized.
\begin{definition} \label{formal1}
Mathematically, a cryptosystem or encryption scheme can be defined as a t-uple $(\mathcal{P}, \mathcal {C}, \mathcal{K}, \mathcal{E})$
where $\mathcal{P}$ is the plaintext space, $\mathcal {C}$ is the ciphertext space, $\mathcal{K}$ is the key space (including keys and subkeys) and 
$\mathcal{E}$ is a set of functions $\{f^k_1, f^k_2, \ldots, g_1, g_2, \ldots\}$ some of them being parametrized by an element $k \in \mathcal{K}$.
\end{definition}
Contrary to Definition in \cite[Section 3.1.1]{Buchmann2004}, we do not make difference between encryption and decryption, symmetric and asymmetric encryption since atomic functions combined are the same. Only the nature of the keys used and/or their order actually defines those particular instances. Without loss of generalities we are now focusing on symmetric encryption and the following definition will make things easier to understand. 

We now give the sets of functions in Definition~\ref{formal1} an internal composition law defined by the composition of functions (provided consistency between functions domain and co-domain; it is however always possible to consider a unique domain/co-domain that provide this consistency). We note $f \circ g = f.g$ for sake of simplicity. To have a unique term we use the term ``\textit{$l$-function word}'' to describe the composition of $l$ functions.
\subsubsection{Formalization of Block Ciphers}
\begin{definition}[Block Ciphers] \label{formal-bc}
Mathematically, a block cipher scheme can be defined as a t-uple $(\mathcal{P}, \mathcal {C}, \mathcal{K}, F_\mathcal{E})$
where $\mathcal{P}$ is the plaintext space, $\mathcal {C}$ is the ciphertext space, $\mathcal{K}$ is the key space (including keys and subkeys) and 
$F_\mathcal{E}$ is the free group of $\mathcal{E} = \{f^k_1, f^k_2, \ldots, g_1, g_2, \ldots\}$ under the function composition law.
\end{definition} 
Thus any block encryption scheme can be described by two words (one for encryption the other for decryption) of $F_\mathcal{E}$. 
\begin{example}
Let us consider the following block ciphers: 
\begin{itemize}
	\item \textbf{Advanced Encryption Standard (AES)} \cite{aes-book}.\\ Let $\mathcal{E} = \{\oplus_{K_i}, S = S\_BOX, R = S\_ROW, M = M\_COL \}$. Then the encryption and decryption functions are described by the words 
	\[w_e = \oplus_{K_{10}}.R.S.\oplus_{K_9}.M.R.B.\oplus_{K_8}\ldots M.R.B.\oplus_{K_0}\]
	and 
	\[w_d = \oplus_{K_{0}}.R.S.\oplus_{K_1}.M.R.B.\oplus_{K_9}\ldots M.R.B.\oplus_{K_{10}}\]
	We can check that $w_e.w_d = Id$.
	\item \textbf{Data Encryption Standard (DES)} \cite[Section 7.4.2]{hac}\\
	Let $\mathcal{E} = \{IP, F_{K_0}, F_{K_1}, \ldots F_{K_{14}}, F_{K_{14}}\}$ where we consider functions on $\mathbb{F}_2^{64}$ (this writing is not unique and we could consider subkey parametrized functions over $\mathbb{F}_2^{32}$ with some rewriting). Hence we have
	\[F_{K_i}(L_i, R_i) = (R_i, L_{i - 1} \oplus f(R_i, K_i))\]
	and 
	\[F_{K_{15}}(L_{15}, R_{15}) = (L_{15} \oplus f(R_{15}, K_{15}), R_{15})\]
	The encryption is then the word $w_e = IP^{-1}.F_{K_{15}}.(\prod_{i = 14}^{0}F_{K_{i}}).IP$
\end{itemize}
Other block ciphers such as Kalyna \cite{kalyna} or Kuznyechik \cite{kuzniechik} can be similarly described.
\end{example}
\subsubsection{Formalization of Stream Ciphers}
\begin{definition}[Stream Ciphers] \label{formal-sc}
Mathematically, a stream cipher scheme can be defined as a t-uple $(\mathcal{K}, \mathcal{E})$
	where $\mathcal{K}$ is the key space and $\mathcal{E} = \{f^k_1, f^k_2, \ldots, g_1, g_2, \ldots\}$ is a monoid equipped with the function composition law.
\end{definition} 
In this definition we do not consider neither plaintext nor ciphertext since stream ciphers, as involutive schemes, apply indifferently on both objects by producing a pseudo-random sequence of finite length $N$ Moreover since functions in $\mathcal{E}$ can be non invertible, it is not possible to consider a more refined algebraic structure. 

It is essential to mention that we can always consider functions having the same domain/co-domain for a simplified description (by basic rewriting using extension or embedding of functions; however it is not mandatory as long as the function composition remains consistent).
\begin{example}
Let us consider the classical case of Linear Feedback Shift Registers (LFSRs) combined by a highly non-linear Boolean function $f$ described in Figure~\ref{fig1}.

Let $\mathcal{E} = \{P_1, P_2, P_3, \wedge_1, b3, f\}$ where the $P_i$ are the feedback polynomials, $f$ is the Boolean combining function over $\mathbb{F}_2^3$, $\wedge_1(x) = x \& 1$ is a Boolean function outputing the least significant bit of an integer and $b3(x_1, x_2, x_3) = (x_3 << 2) | (x_2 << 1) | x_1$ computes a 3-bit value from three signgle bits. Then to produce a sequence $(\sigma_t)_{1 \leq t \leq N}$ of length $N$ we note
\[\sigma_t = f.b3.(\wedge_1(P^t_1), \wedge_1(P^t_2), \wedge_1(P^t_3)))\] 
\end{example}
 \subsection{Combinatorial Equivalence and Cryptanalysis}
 Now we have defined (symmetric) cryptosystem conveniently, we intend to make the concept of \textit{Combinatorial Equivalence} evolve. Initially it has been introduced in combinatorial group theory \cite{combgt,Newman42} along with the concept of Tietze transformations \cite{tietze}. 
 \begin{definition}\cite{Newman42}
 \textit{Combinatorial Equivalence} is an equivalence relation between object defined by means of certain ``allowed transformations''. Two objects are regarded as ``combinatorially equivalent'' if and only if one is obtainable from the other by a series of allowed transformations.
 \end{definition}  
However this definition is used to transform a given presentation of a group into another, often simpler presentation of the same group.  Transformations are built up of elementary steps, each of which individually takes the presentation to a presentation of an isomorphic group. In our case this property of isomorphism is too limitative. 

Our idea is to modify the set of functions $\mathcal{E}$ in order to produce a combinatorially equivalent cryptosystem that make some unsuspected properties/structure appears that give a stronger information linked to the secret used. 

It is expected that this ``equivalent'' enables a less complex cryptanalysis. It is essential to stress on the fact that we do not think that this could enable to change the complexity class of the underlying problem to be solved to break a cryptosystem. But we think however that it is possible to switch to more tractable instances in the same complexity class.
 The general approach is described by the following diagram 
 \begin{equation} \label{diag1}
 	\begin{CD}
 		\mathcal{P} @>{\tau}>>\mathcal{P}_{\tau}\\
 		 @VV\operatorname{w_e}V         @VV\operatorname{w'_e}V\\
 		\mathcal{C} @>\tau'>> \mathcal{C}_{\tau'}
 	\end{CD}
 \end{equation}
A number of conditions are fixed:
\begin{itemize}
\item Functions $\tau$ and $\tau'$ are at least homomorphism. The isomorphism property is not mandatory.
\item $w_e$ and $w'_e$ are respectively $l$- and $l'$-function words but generally $l < l'$. In other words, related sets $\mathcal{E}$ and $\mathcal{E'}$ may be (totally or not) different and hence may be of different order. For instance, in \cite{bea1}, partition-related functions have been added to the usual set of functions $\mathcal{E}$ used in block ciphers. 
\item Diagram~\ref{diag1} commutes. It means that we have $\tau'.w_e = w'_e.\tau$.
\end{itemize}   
The very final aim is that $w'_e$ introduces properties with respect to the secret key $K$ that are probabilistically stronger than with $w_e$. 
\section{Stream Cipher Cryptanalysis State-of-the-Art} \label{sec:sota}
While there is a large literature dealing with stream ciphers based on LFSRs and their cryptanalysis, there is quite no result as far as their non-linear analogues are concerned.

Stream cipher cryptanalysis techniques are essentially divided in two categories: statistical/correlation attacks and algebraic attacks. Other variants fall into one or both of these categories.
\subsection{Statistical Correlation Attacks}
\subsubsection{Simple Correlation Attacks}
All these attacks rely on a sufficiently high correlation between the sum of a subset of the input variables and the output of the combining Boolean function $f$ \cite{Sieg1985}. 
In other words $p = P[f(x) = <x, u>]$ where $<.,.>$ denotes the bitwise scalar product. The Hamming weight of $u$ (which describes the subset of input registers to be taken into account) determines how many registers have to be considered to go through an exhaustive search simultaneously. 

Considering a known plaintext attack (as long as this is operationally acceptable), the required keystream length $N$ depends on the probability on the correlation probability $p_i$, on the length $L$ of the register(s), on the probability of false alarm $P_f$ (\textit{i.e.} the probability that a wrong key is kept as a good candidate) and on the probability of non detection $P_m$ (the good candidate has been rejected and is missing). With $P_m = 10^{-3}$ and $P_f = 2^{-L}$ we then have \cite{canteaut2011} 
\begin{equation} \label{eq1}
	N < \Bigg(\frac{\sqrt{L} + 3\sqrt{2.p.(1 - p)}}{\sqrt{2}(p - 0.5)}\Bigg)^2
\end{equation}
From Equation~\ref{eq1}, it follows that the required data size $N$ increases inversely with the probability $p$ and increases with the register length $L$.

Correlation attacks apply to both LFSR-based and NLFSR-based combining stream ciphers.
\subsubsection{Fast Correlation Attacks}
Fast correlation attacks \cite{FastCorr1988,FastCorr1989} rely on the same principle as the correlation attack since they exploit the existence of a correlation between the sum of a subset of the input variables and the output of the combining Boolean function $f$. But instead of searching through all the possible initializations of target LFSRs, they model LFSR sequences as error-correcting codes on which the combining Boolean function operates as noise of parameter $p$ (Binary Symmetric Channel). The key recovery step consists then in applying a maximum-likelihood decoding algorithm to the linear code defined
by the LFSR(s) feedback polynomial(s).

While they are faster than correlation attacks, they require far larger input data (output bits from $f$ or ciphertext bits). In this respect, the effectiveness of fast correlation attacks can be questioned in most operational use-cases. A exhaustive survey on fast correlation attack techniques can be found in \cite{Joensson2002}.

It is worth mentioning that fast correlation attacks regarding NLFSRs-based combiners do not apply. Up to the authors' knowledge, no study has been published on the possible generalization to these combiners.
\subsection{Algebraic Attacks}
Algebraic attacks~\cite{ac2003} and their fast version~\cite{fac2003} consist in expressing stream ciphers output bits as equations of low degree where the unknowns are  the initialisation bits of the linear shift registers.

The general principle of algebraic attacks is to recover the key (the registers' initialisations) by solving the system of such equations (in a known plaintext context). Generally the number of such equations can be much larger than the number of unknowns. This makes the resolution of the
system less complex. The resolution is performed either by using Groebner bases~\cite{groebner2009} or by linearizing the system (replacing every monomial of degree greater
than 1 by a new unknown variable). The resulting linear system has however too many unknowns (especially is the degree of the equation is beyond 2 or 3) in practice and cannot be solved whenever the algebraic degree of the combining function is large enough. The complexity of these attacks and the required consecutive output bits make them not effective in practice.

These attacks do not apply at the present time to NLFSRs combiners.
\section{CipherBent6 Specifications} \label{sec:algospecs}
\begin{definition}[Concept Cipher]
	A \textit{concept cipher} $\texttt{E}$ is a cryptographic algorithm describing a family $\mathcal{A}$ of cryptographic algorithms of which it captures all the complexity. It then allows the cryptographic security analysis of $\mathcal{A}$ from its representative $\texttt{E}$. We then can state: 
	\begin{itemize}
		\item An effective cryptanalysis on $\texttt{E}$ can be verified by challenge or demonstration, effectively.
		\item Any cryptanalysis that would work on any member of $\mathcal{A}$ would also be efficient on $\texttt{E}$ with a lower complexity. 
		\item A reliable evaluation of the cryptanalysis of systems of the same class can be derived from an effective cryptanalysis of $\texttt{E}$. As a consequence, any applicable cryptanalysis on $\texttt{E}$ allows to give an evaluation of the cryptanalysis complexity for the algorithms of $\mathcal{A}$.
		\item Any cryptanalysis of $\texttt{E}$ is transposable/scalable to any system of class $\mathcal{A}$.
	\end{itemize}
	The concept cipher $\texttt{E}$ is thus a minimal element of $\mathcal{A}$ ordered by the order relation with respect to the cryptanalysis complexity.  
\end{definition}
In  this paper, we consider a concept cipher called \textit{CipherBent6}. The corresponding class is that of stream ciphers in which Non-Linear Shift Registers (NLFSR) are combined by a non-linear Boolean function. Achterbahn Algorithm \cite{Acht2006} is a member of this class. Among many other aspects,
concept Cipher $\texttt{E}$ is worth considering whenever it resists to all known attacks.

As for \textit{CipherBent6}, the combining function is 6-variable bent function.
Its main specifications are the following:
\begin{itemize}
	\item The combined function is a 6-variable bent function $x_1x_4 \oplus x_2x_6 \oplus x_1x_4$ \cite{Rot76}. It has maximal non-linearity $\mathcal{NL}(f) = 28$. Its algebraic immunity is not optimal since $\mathcal{AI}(f) = 2 < 3$. However, this non-optimal nature is not exploitable in the case of non-linear feedback shift registers.
	\item Six nonlinear linear shift registers (NLFSR) of respective length $27, 28, 30, 31,$ $32, 33$, all having maximum period. These NLFSRs are those used in the Achterbahn stream cipher \cite{Acht2006}. Their feedback polynomial are all dense and of high degree. For instance, NLFSR $A_{12}$ has the following Algebraic Normal Form (ANF) for its feedback polynomial:
	\small
 \begin{equation}\begin{split}
		A_{12}(x_0,\ldots, x_{32}) &= \\& x_0 \oplus x_2 \oplus x_7 \oplus x_9 \oplus x_{10} \oplus x_{15} \oplus x_{23} \oplus x_{25} \oplus x_{30} \oplus x_8x_{15} \oplus x_{12}x_{16}\\
		& \oplus x_{13}x_{15} \oplus x_{13}x_{25} \oplus x_1x_8x_{14} \oplus x_1x_8x_{18} \oplus x_8x_{12}x_{16} \oplus x_8x_{14}x_{18}\\
		& \oplus x_8x_{15}x_{16} \oplus x_8x_{15}x_{17} \oplus x_{15}x_{17}x_{24} \oplus x_1x_8x_{14}x_{17} \oplus x_1x_8x_{17}x_{18}\\
		& \oplus x_1x_{14}x_{17}x_{24} \oplus x_1x_{17}x_{18}x_{24} \oplus x_8x_{12}x_{16}x_{17} \oplus x_8x_{14}x_{17}x_{18} \\
		& \oplus x_8x_{15}x_{16}x_{17} \oplus x_{12}x_{16}x_{17}x_{24} \oplus x_{14}x_{17}x_{18}x_{24} \oplus x_{15}x_{16}x_{17}x_{24}.
   \end{split}\end{equation} \normalsize
	\item The key is the initial content of the 6 registers at time $t = 0$. Thus the key is 181 bit long.
\end{itemize}
\section{Cipherbent6 Cryptanalysis Results} \label{sec:formal}
Cipherbent6 has been designed as a concept cipher for non-linear shift register-based combiners and to test our CE-cryptanalysis. \textbf{Cipherbent6 is in no way intended to be a secure encryption system nor usable for real applications except as an evaluation/concept cipher}.

The aim of our work is to provide a cryptanalysis of Cipherbent6 which is as much effective as possible. From a limited size output (known plaintext attack), we intend to recover the 181-bit key. 

When considering the existing attacks against stream ciphers presented in Section~\ref{sec:sota}, none of them are applicable except correlation attacks but with an overall complexity of $2^{35}$ provided that at least 8,000 output bits are known (using Equation~\ref{eq1} and the fact that for a 6-variable bent function $p = 0.5625$). However, our own simulation results show than in practice around 10,000 bits are necessary to recover the 181-bit key uniquely (with a negligible amount of wrong candidates). As for other techniques, their inapplicability comes mostly from the fact that we deal with dense NLFSRs combined with a complex Boolean function. 

The cryptanalysis of Cipherbent6 is in two parts:
\begin{enumerate}
	\item \textbf{Research part}.- Obtaining combinatorial equivalents of Cipherbent6 is a \textbf{one-time operation} that takes from 24 hours (version 1) to 48 hours (version 2) and has overall complexity of $\mathcal{O}(2^{43})$. This step is independent from the initialisation of the registers (which actually is the secret key). At the present time these two first combinatorial equivalents we have produced may not the most optimal ones. Our initial goal was first to find at least one equivalent to validate our approach and second to prove that more optimal equivalents do exist. 
	
	\item \textbf{Exploitation part}.- From a given combinatorial equivalent, the cryptanalysis part requires $N = 2,790$ output bits (version 1) or $N = 1,820$ output bits (version 2). For both versions, it requires a computing time of slightly less 72 hours to retrieve the 181 bits of key. The overall complexity is in $\mathcal{O}(2^{45})$. These preliminary results outperform the known attacks that could be applied (basically correlation attack). This cryptanalysis is not optimized yet and we expect to reduce the cryptanalysis time as well. 
\end{enumerate}

Interested people can submit a challenge for us to solve by visiting the \textit{Hope4Sec} website (where the reference source code for Cipherbent6 is provided).

Experiments have been conducted on two AMD Ryzen Threadripper 2990 WX 32-Core Processor x64 (256 Mb of RAM, 36 Tb HDD each). 

\section{Conclusion et future work} \label{sec:conc}
We are currently working to identify more efficient combinatorial equivalents of Cipherbent6. Different candidates are currently under analysis. We expect to reduce the number of the output bits required to less than 1,000 bits in the next step. The ultimate goal/hope is to achieve a number of bits close to the size of the secret key.

The next step is to apply the CE technique to real encryption systems. The first candidate that is closest to Cipherbent6 is the \textit{Achterbahn} algorithm. We expect to find an effective cryptanalysis with around a few Kilobits of output bits (less than 500 Kb).

Finally, our current research deals with the application of CE cryptanalysis to block ciphers (especially the Russian cryptographic standard Kuznyechik \cite{kuzniechik} cipher). We already have identified combinatorially equivalent primitives with stronger probabilistic biases that could be used to build function word $w'_e$.


\begin{credits}
\subsubsection{\discintname}
There are no ethical issues. The authors do not have any competing interest of any kind. This research work was entirely self-financed and is the exclusive research work of the authors.
\end{credits}

\begin{flushright}
	\textbf{S. D. G.}
\end{flushright}
%
%
%
%

\end{document}